\documentclass[aps,prl,twocolumn,superscriptaddress]{revtex4-1}

\usepackage{graphicx}
\usepackage{amsmath}
\usepackage{mathtools}
\usepackage{braket}

\begin{document}

\title{Observation of dipole-quadrupole interaction in an ultracold gas of Rydberg atoms}

\author{Johannes Deiglmayr}
\affiliation{Laboratory of Physical Chemistry, ETH Zurich, Vladimir-Prelog-Weg 2, 8057 Zurich, Switzerland}
\email{jdeiglma@ethz.ch}
\author{Heiner Sa{\ss}mannshausen}
\affiliation{Laboratory of Physical Chemistry, ETH Zurich, Vladimir-Prelog-Weg 2, 8057 Zurich, Switzerland}
\author{Pierre Pillet}
\affiliation{Laboratoire Aim\'e Cotton, CNRS, Univ Paris-Sud 11, ENS-Cachan, Campus d'Orsay Batiment 505, 91405 Orsay, France}
\author{Fr\'ed\'eric Merkt}
\affiliation{Laboratory of Physical Chemistry, ETH Zurich, Vladimir-Prelog-Weg 2, 8057 Zurich, Switzerland}

\newcommand{\AR}{{\textrm{A}}}
\newcommand{\SF}{{$ns\,n'f$}}
\newcommand{\PP}{{$np\,np$}}
\newcommand{\BR}{{\textrm{B}}}

\begin{abstract}
We observe the direct excitation of pairs of Cs atoms from the ground state to molecular states correlating asymptotically to \SF\ asymptotes. The molecular resonances are interpreted as originating from the dipole-quadrupole interaction between the \SF\ pair states and close-by \PP\ asymptotes ($22\leq n \leq 32$). This interpretation is supported by Stark spectroscopy of the pair states and a detailed modeling of the interaction potentials. The dipole-quadrupole interaction mixes electronic states of opposite parity and thus requires a coupling between electronic and nuclear motion to conserve the total parity of the system. This non-Born-Oppenheimer coupling is facilitated by the near-degeneracy of even and odd $L$ partial waves in the atom-atom scattering which have opposite parity.
\end{abstract}

\maketitle

The excitation of ultracold atoms to interacting Rydberg states has given rise to the observation of a wide range of fundamental phenomena such as the formation of ultracold neutral plasmas~\cite{robinson2000,killian2007}, the conditional blockade of excitation (dipole-blockade) that can be exploited to realize basic quantum gates~\cite{isenhower2010,wilk2010}, and the formation of macrodimers~\cite{farooqi2003,overstreet2009}. In a macrodimer, two atoms are bound by long-range dispersion forces between two highly excited Rydberg atoms~\cite{boisseau2002}. Most observations could be interpreted as arising from dipole-dipole interactions, but occasionally quadrupole-quadrupole interactions have been found to be relevant~\cite{stanojevic2008,schwettmann2006}. These interactions conserve the electronic parity of the system of two interacting Rydberg atoms and obey the condition
\begin{equation}\label{eq:parCond}
(-1)^{\ell_{\AR,i}+\ell_{\BR,i}}=(-1)^{\ell_{\AR,f}+\ell_{\BR,f}},
\end{equation}
where $\ell$ represents the orbital angular momentum quantum number of the Rydberg electron, the indices $i$ and $f$ the initial and final pair states coupled by the interaction, and $\AR$ and $\BR$ the two atoms. Typical examples include $ss$-$pp'$ interactions (using the notation $\ell_{\AR,i}\ell_{\BR,i}$-$\ell_{\AR,f}\ell_{\BR,f}$), which are exploited in dipole-blockade experiments~\cite{walker2008}.

We report on the observation of sharp $pp$-$sf$ pair resonances in the Rydberg spectrum of a dense ultracold Cs atom sample following excitation with an intense narrow-band pulsed UV laser. $pp$-$sf$ pair resonances violate condition~\eqref{eq:parCond} and are attributed to dipole-quadrupole interactions between pairs of Rydberg atoms. The surprising observation is the strength and ubiquity of these resonances in the spectrum of Cs, which, in retrospect, makes it surprising that they have not been observed or discussed before. Dipole-quadrupole interactions violate some of the symmetries usually employed to designate the quantum states of a homonuclear diatomic molecule. These symmetries include ($i$) the inversion symmetry, or g/u symmetry, of the electronic wavefunction upon inversion through the symmetry center, ($ii$) the electronic parity, or $\pm$ electronic symmetry, of the electronic wavefunction upon inversion of all coordinates through an arbitrary point of the space-fixed coordinate system - this operator is formally equivalent to a reflection $\sigma_v$ of the electronic wavefunction through a plane containing the internuclear axis~\cite{hougen1970,lefebvre2004} - ($iii$) the total (rovibronic) parity, and ($iv$) the total angular momentum $\vec{J}$ (or $\vec{F}$ if nuclear spins are considered) and its projections $\hbar M$ on a chosen space-fixed axis and $\hbar \Omega$ on the internuclear axis. In free space and neglecting the electroweak force~\cite{quack2011}, only ($iii$) and ($iv$) are strictly conserved. Because the total parity can be obtained as a product of the parities of the electronic and rotational wavefunctions, the violation of condition~\eqref{eq:parCond} observed in our experiments implies the breakdown of the $\pm$ and g/u symmetries and the resulting entanglement of rotational and electronic motions.

In the standard treatment of long-range interactions between two particles $\AR$ and $\BR$~\cite{hirschfelder1967,dalgarno1967} and its recent application to the treatment of Rydberg pair resonances~\cite{flannery2005,farooqi2003,overstreet2007}, the interaction term $V_\textrm{inter}$ of the total Hamiltonian (in atomic units)
\begin{align}\label{eq:Htot}
H &=H^\AR_0+H^\BR_0+V_\textrm{inter}  \\
V_\textrm{inter} & =  \sum_{\mathclap{L_{\AR/\BR}=1}\hspace{0.5cm}}^\infty \sum_{\hspace{0.5cm}\mathclap{\Omega=-L_<}}^{+L_<}\frac{(-1)^{L_\BR}f_{L_\AR L_\BR \Omega}}{R^{L_\AR+L_\BR+1}} Q_{L_\AR \Omega}(\vec{r}_\AR)Q_{L_\BR -\Omega}(\vec{r}_\BR) \nonumber \\
Q_{L\Omega}(\vec{r}) &= \sqrt{\frac{4 \pi}{2 L +1}} r^L Y_{L\Omega}(\hat{r}) \nonumber \\
f_{L_\AR L_\BR \Omega} & =  \frac{(L_\AR+L_\BR)!}{\sqrt{(L_\AR+\Omega)!(L_\AR-\Omega)!(L_\BR+\Omega)!(L_\BR-\Omega)!}}    \nonumber
\end{align}
is expressed in the frame of the nonrotating molecule $\AR\BR$~\cite{fontana1961}, in order to obtain potential-energy functions, which adequately describe the interaction at long range. Eq.~\eqref{eq:Htot} therefore neglects, by design, electronic-rotational interactions. In Eq.~\eqref{eq:Htot}, $H^i_0$ is the unperturbed Hamilton operator of atom $i$, $Q_{L_i\Omega}(\vec{r}_i)$ corresponds to the $L_i$-th multipole moment of atom $i$ ($L_i$=1 and $L_i$=2 are the dipole and the quadrupole moment, respectively), and $L_<$ is the smaller value of $L_\AR$ and $L_\BR$. The different terms of the sum in $V_\textrm{inter}$ transform under $\sigma_v$ as $(-1)^{L_\AR+L_\BR}$~\footnote{The $Y_{L\Omega}$ functions transform as $(-1)^{L-\Omega}$ under $\sigma_v$~\cite{hougen1970}, regardless  of the position of their center along the internuclear axis.}, which implies that only multipole interactions with even values of $L_\AR+L_\BR$ conserve the electronic parity.

The experiments were performed on Cs atoms released from a crossed optical dipole trap ($\lambda=1064$~nm, $P=10$~Watt). Using saturated absorption imaging, we determine typical particle numbers of $N=10^6$, densities of $n=10^{12}\,\textrm{cm}^{-3}$, and temperatures of 40~$\mu$K. We employ a one-photon transition to excite atoms from the 6$s_{1/2}$ ground state to $np$ Rydberg states. The UV light ($\lambda\approx320$\,nm) is provided by a cw ring dye laser (Coherent 899-21) which is frequency doubled after pulsed amplification~\cite{salour1977}. The laser pulses are 4.4\,ns long with a transform-limited band width of 140~MHz and pulse energies of up to 100~$\mu$J. The laser is focused down to a waist of $\sim$\,100~$\mu$m, still significantly larger than the size of the atomic sample ($\sigma\approx 30$~$\mu$m). Varying the intensity of the UV laser did not lead to an observable shift of the transitions. The laser frequency is calibrated using a wavemeter (High Finesse WS6-200). For detection, we rely on spontaneous ionization of the excited Rydberg sample by photoinitiated collisions (PIC)~\cite{overstreet2007} and subsequent detection of the resulting ions. Alternatively, we can detect $np$ states with $n \ge 27$ using pulsed field ionization (PFI)~\cite{sasmannshausen2013}. Spectra obtained with PIC and PFI are very similar, however the former have a better signal-to-noise ratio and thus only those are shown.

\begin{figure}
\begin{center}
\includegraphics[width=0.99\linewidth]{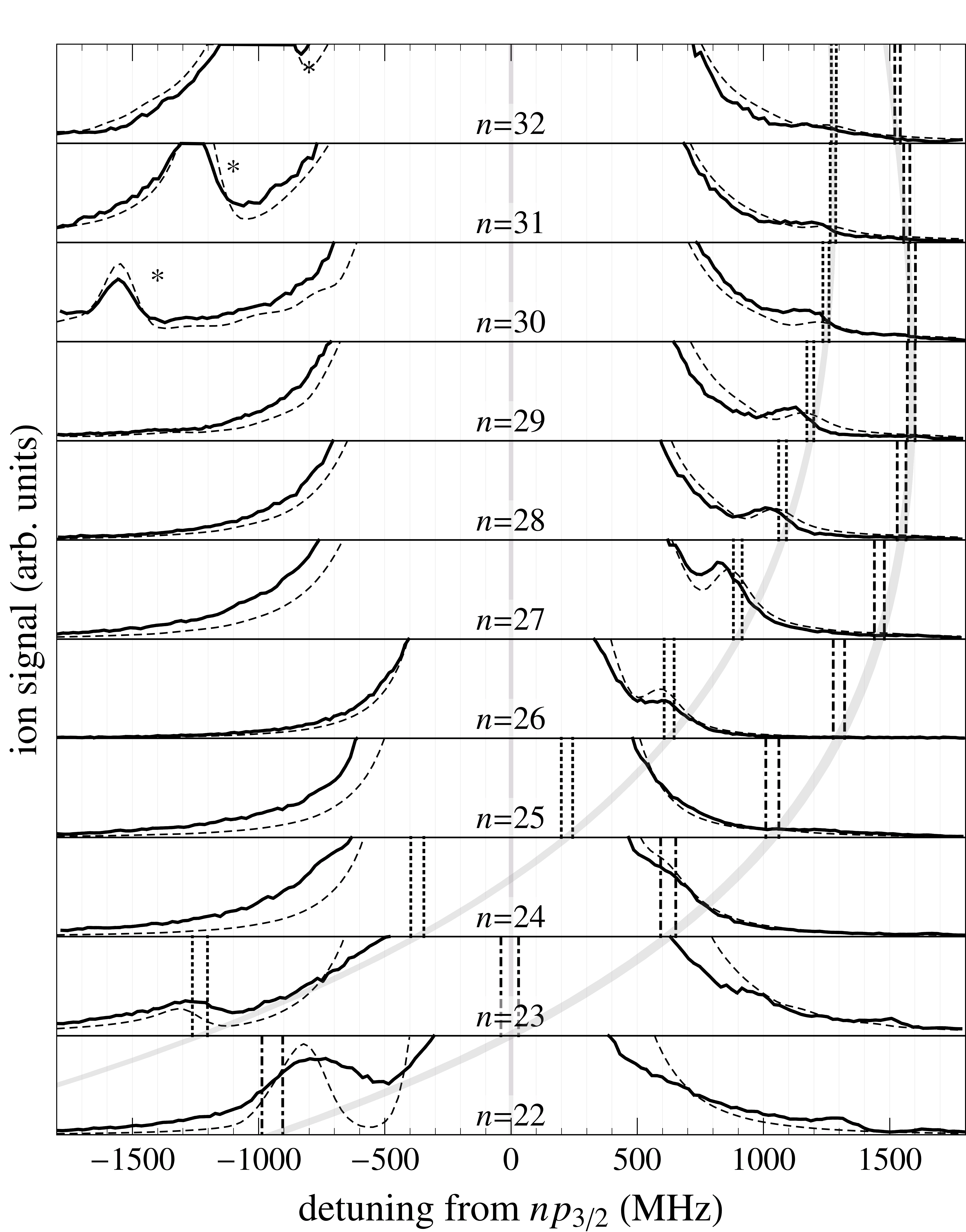}
\caption{Comparison of experimental spectra (thick lines) with simulated line profiles (dashed lines). The asymptotic pair energies of the $ns_{1/2}(n-3)f_J$ (J=$5/2,7/2$) and $(n+1)s_{1/2}(n-4)f_J$ states are indicated by vertical dotted and dashed-and-dotted lines, respectively. Experimental and simulated spectra have been scaled separately for each $n$ to optimize the visibility of spectral features. The $n$-dependence of the asymptotic pair energies is visualized by gray bands connecting their positions. The features in the traces for $30\le n\le 32$  marked by asterisks originate from $ns(n+1)s$ pair states.}\label{fig:allspec}
\end{center}
\end{figure}

The dipole-quadrupole interaction is observed as the appearance of additional resonances in Rydberg spectra around $np_{3/2}$ $ (22 \le n \le 32$), see Fig.~\ref{fig:allspec}, when the Rydberg-atom density is increased by increasing the UV laser intensity. The positions of these resonances correspond closely to the asymptotic energies of pairs of Rydberg atoms in $ns_{1/2}(n-3)f_J$ ($J$=$5/2,7/2$) and $(n+1)s_{1/2}(n-4)f_J$ states indicated in each spectrum by vertical lines. For $n<22$, larger detunings and reduced matrix elements cause the \SF\ pair resonances to be too weak to be observable in our experiment. For $n>32$, more strongly coupled pair resonances mask the \SF\ resonances. At $n=23$, the two fine-structure components of the $24s_{1/2}18f_J$ pair state are almost degenerate with the $23p_{3/2}23p_{3/2}$ pair state, and at $n=24,25$, the detunings of the $ns_{1/2}(n-3)f_J$ pair states from the $np_{3/2}np_{3/2}$ pair states are too small to be resolved in our experiment.

\begin{figure}
\begin{center}
\includegraphics[width=1.0\linewidth]{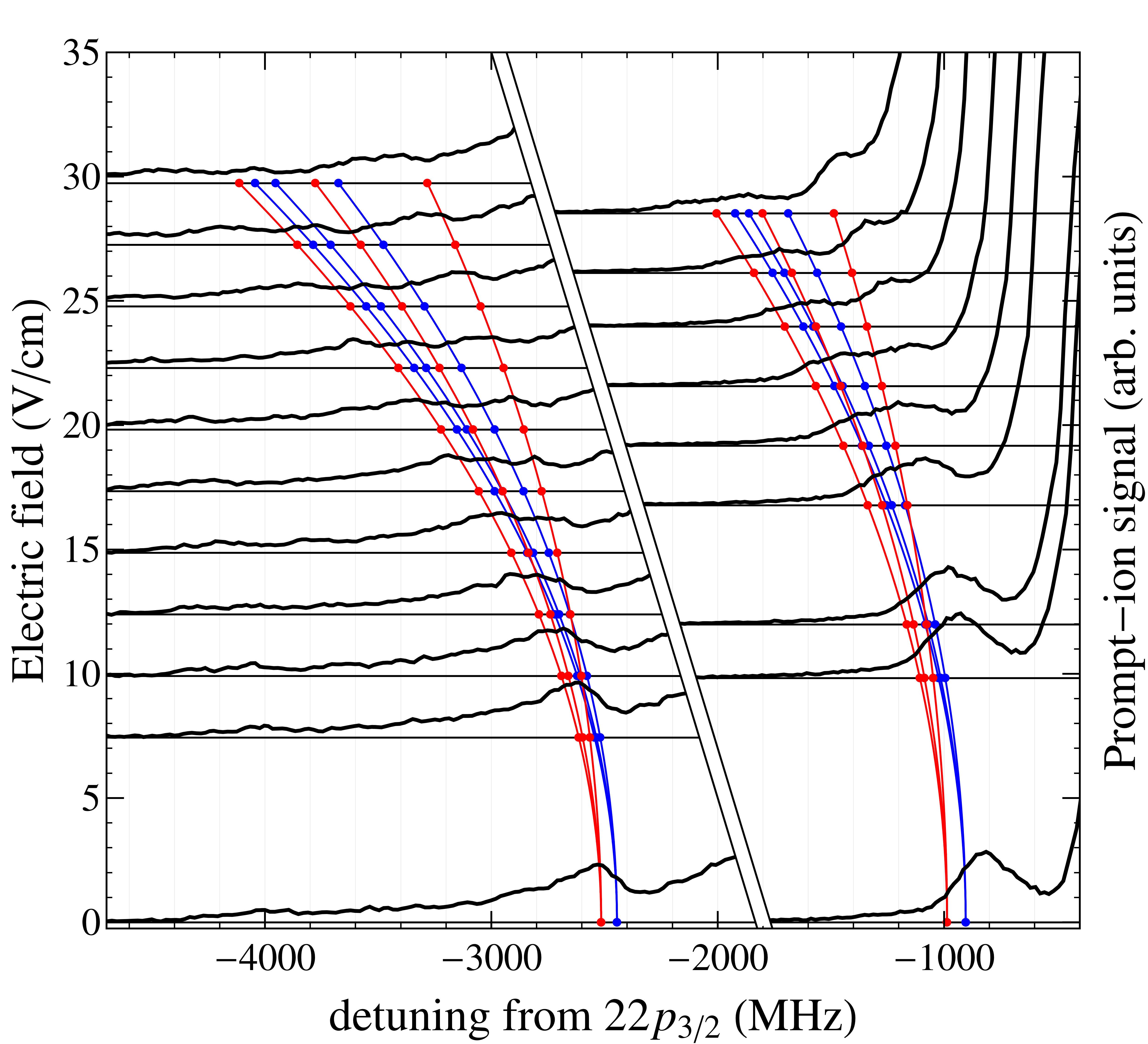}
\caption{(color online) Experimental spectra of the $22s_{1/2}19f_J$ and $23s_{1/2}18f_J$ pair states at different electric fields (left and right black traces respectively) and corresponding calculated positions of the atomic asymptotes (blue and red curves corresponding to $s_{1/2}f_{5/2}$ and $s_{1/2}f_{7/2}$ respectively). The experimental spectra are offset by the respective electric field value in V/cm (marked by the solid base line of each spectrum).}\label{fig:starkspec22}
\end{center}
\end{figure}

The excitation of \SF\ pair states could in principle also result from $\ell$ mixing induced by an external electric field~\cite{schwettmann2006}. Using Stark spectroscopy of high-lying Rydberg states, we estimate residual stray electric fields to be less than 50~mV/cm. For the investigated quantum numbers $22\le n\le32$ such small electric fields lead to negligible $\ell$ mixing. We experimentally verified this expectation by intentionally applying larger electric fields and observing the shift of the \SF\ pair resonances (see Fig.~\ref{fig:starkspec22} for a measurement at $n=22$). The observed line shifts are in good agreement with the calculated asymptotic Stark shifts of the pair states, which are the sum of the Stark shifts of the corresponding atomic states, calculated following Ref.~\cite{zimmerman1979}. Although the spectral features are broadened because of the splitting of the magnetic sublevels by the field, the integrated line intensity remains approximately constant. If the observation of \SF\ pair resonances were caused by an external electric field, one would instead expect an increase in the line intensities by several orders of magnitude.

To model the experimental spectra, we determine the matrix elements of the electronic Hamiltonian~\eqref{eq:Htot} for $L_\AR$,$L_\BR$=1,2 and fixed $\Omega$=$\omega_\AR$+$\omega_\BR$ in the atomic basis $\ket{n_\AR \ell_\AR j_\AR \omega_\AR , n_\BR \ell_\BR j_\BR \omega_\BR}\equiv\ket{ \gamma_\AR j_\AR \omega_\AR, \gamma_\BR j_\BR \omega_\BR}$, following the general approach of Refs.~\cite{stanojevic2008,stanojevic2006,samboy2011}. The Hamiltonian omits interatomic electron-correlation. This omission is permitted if the internuclear distance exceeds the LeRoy-radius $R_\textrm{LR}=2\left(\braket{r_\AR^2}^{1/2}+\braket{r_\BR^2}^{1/2}\right)$~\cite{leroy1973}, which is the case for all distances considered in this work. Retardation effects can also be neglected, because the reduced wavelengths of the relevant Rydberg-Rydberg transitions are much larger than the distances~\cite{hirschfelder1967} considered here. The hyperfine splittings of the relevant Rydberg states have been investigated in a previous work~\cite{sasmannshausen2013} and are much smaller than any other energy scale in the system, justifying their omission. Radial matrix elements are obtained by numerical integration using experimental quantum defects and an \textit{ab-initio} model potential~\cite{zimmerman1979,goy1982,marinescu1994}.

From the symmetry argument presented above, the $\pm$ electronic symmetry mixing of the dipole-quadrupole contribution must be compensated by a change of the rotational parity or the parity $(-1)^L$ of the collisional partial wave with angular momentum $\hbar L$, as discussed for collision-induced transitions between rotational levels of ground-state molecules~\cite{oka1974}. Considering the pair interaction as a classical collision with reduced mass $\mu$, occurring at distance $\overline{R}$ with relative velocity $v=\sqrt{{k_\textrm{\BR} T}/{m_\textrm{Cs}}}$ and maximal impact parameter $b=\overline{R}$, the maximum angular momentum of the collision complex is given by $\hbar L_\textrm{max}=\mu \overline{R} v$. The energy splitting between rotational states with angular momentum $L$-1 and $L$ is $\Delta E_\textrm{rot}=2 L B_\textrm{rot}=L\frac{\hbar^2}{\mu \overline{R}^2}$. The maximum rotational energy splitting between dipole-coupled rotational states ($\Delta L =\pm1$) is thus $\Delta E_\textrm{rot,max}=\hbar{v}/\overline{R}$. For the distances relevant to this work (\textit{e.g.}, $\overline{R} \approx 0.23$~$\mu$m for $n=22$, see Fig. 3~a)) and ultracold temperatures, $\Delta E_\textrm{rot,max}$ is always smaller than the linewidth of the pair states (approximated by the sum of atomic linewidths from Ref.~\cite{ovsiannikov2011}) and therefore the rotational states of opposite parity can be considered as degenerate. We therefore assume the coupling between electronic and rotational motion to be the same for interaction terms which require a change in $L$ (\textit{i.e.}, dipole-quadrupole) as for terms which do not require a change in $L$ (\textit{i.e.}, dipole-dipole and quadrupole-quadrupole). This assumption is justified \textit{a posteriori} by the agreement between simulated and experimental spectra, as is now explained.

\begin{figure}
\begin{center}
\includegraphics[width=1.0\linewidth]{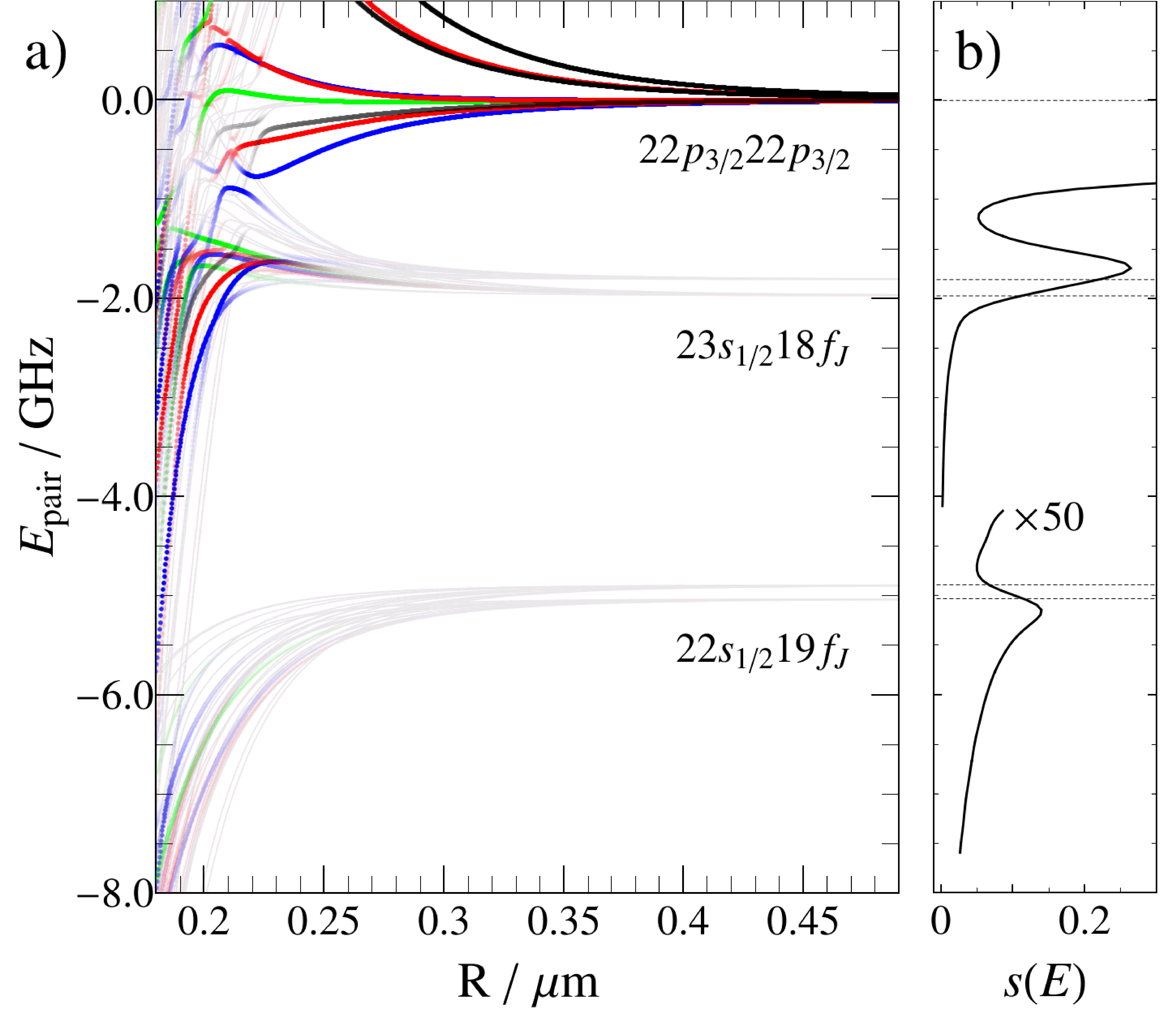}
\caption{(color online) a) Calculated potential-energy curves for $n=22$ and  $\Omega=0,1,2,3$ (black, red, blue, and green curves respectively). The intensity of the color codifies the value of  $\overline{p_\Psi}$ as defined in the text (gray corresponding to zero, full color to more than 5\% overlap with the excited atomic state). b) Simulated excitation profile using Eq.~\eqref{eq:spec}. Dotted lines indicate the asymptotic pair energies.}
\label{fig:theocal22}
\end{center}
\end{figure}

Diagonalization of $H$ for different internuclear separations $R$ yields the eigenstates $\ket{\Psi(R)}=\sum \alpha_{\gamma_\AR j_\AR  \omega_\AR, \gamma_\BR j_\BR \omega_\BR }(R)\ket{\gamma_\AR  j_\AR \omega_\AR, \gamma_\BR j_\BR \omega_\BR }$ with eigenenergies $E_\Psi(R)$~\footnote{In order to reproduce the experimental spectrum around the asymptote $np\,np$ we found it necessary to include the pair states $n_\AR s$-$n_\BR s$, $n_\AR p$-$n_\BR p$, $n_\AR d$-$n_\BR d$, $n_\AR f$-$n_\BR f$, $n_\AR s$-$n_\BR f$, and $n_\AR p$-$n_\BR d$ with $|n_{\AR/\BR}^* - n_p^*|\le 1$ and $|E_{n_\AR \ell_\AR,n_\BR \ell_\BR}-E_{n_p pn_p p}|\le 0.6 |E_{n_p pn_p p}-E_{(n_p-1) p (n_p-1) p}|$ in the basis set.}. The resulting potential-energy curves are shown in Fig.~\ref{fig:theocal22}\,a) for $n=22$ (gray lines). The states correlating to the $22s_{1/2}19f_J$ asymptotes are all attractive, while the states correlating to the $23s_{1/2}18f_J$ asymptotes are all repulsive. The origin of this apparent repulsion between these asymptotes, which are not directly coupled by $H$, lies in the large number of coupled states with both negative and positive detuning, especially the dipole-dipole coupled asymptotes $20d23p$ and $21d22p$ at detunings of approximately $-60$~GHz and +100~GHz, respectively. The coupling of pair states with $pp$ and $sf$ character leads also to the appearance of minima in some potential energy curves. In Fig.~\ref{fig:theocal22}~a) such a minimum for a state with $\Omega=2$ (blue curves) is visible below the $22p_{3/2}22p_{3/2}$ asymptote. This state has a strong $pp$ character and represents a bound macrodimer under field-free conditions. We estimate a vibrational frequency of about 20~MHz, which is larger than the expected linewidth and we predict that the vibrational structure of this state should be observable.

The direct diagonalization of $H$ in the atomic basis results in a large number of potential-energy curves. However, very few of them can be excited from the $6s_{1/2}6s_{1/2}$ ground state. The excitation probability $p_\Psi(R)$ of an eigenstate $\ket{\Psi}$ is proportional to the square of the overlap of the molecular wavefunction with the laser-excited $np_{3/2}np_{3/2}$ pair state $\ket{\phi}=\sum a_{j_\AR m_\AR, j_\BR m_\BR}\ket{j_\AR m_\AR, j_\BR m_\BR}$ in the laboratory-fixed system
\begin{align*}
p_\Psi(\theta,R)&=\left|\braket{\Psi(R)|\phi}\right|^2=\\&\Big|\sum_{\mathclap{\substack{\gamma_\AR j_\AR  \omega_\AR, \gamma_\BR j_\BR \omega_\BR  \\
j_\AR m_\AR, j_\BR m_\BR}}}\braket{\Psi(R)|\gamma_\AR j_\AR  \omega_\AR, \gamma_\BR j_\BR \omega_\BR } d_{\omega_{\AR} m_{\AR}}^{j_{\AR}}(\theta) \times\\[-0.6cm] & \hspace{2.8cm} d_{\omega_{\BR} m_{\BR}}^{j_{\BR}}(\theta) \braket{j_\AR m_\AR, j_\BR m_\BR|\phi}\Big|^2 ,
\end{align*}
where $\theta$ is the angle between the internuclear and the laboratory-fixed $z$ axis and $d_{\omega m}^j(\theta)$ is Wigner's $d$ matrix describing the rotation from the laboratory-fixed system to the molecular-fixed system~\footnote{Because of a Cooper-minimum in the photoexcitation cross section from the 6$s_{1/2}$ state to the $np_{1/2}$ ($\epsilon p_{1/2}$) states (continuum) of Cs just above the ionization threshold~\cite{raimond1978}, only the photoexcitation to $np_{3/2}np_{3/2}$ states is of relevance.}. Neglecting nuclear spin and assuming an initial random population of the $m_\AR$ and $m_\BR$ components of the $6s_{1/2}$ ground state, one obtains four possible pair states $\ket{\phi_i}$ after excitation with circularly polarized laser radiation. We obtain $p_\Psi$ by summing over these four initial states with equal weights. After averaging $p_\Psi(\theta,R)$ over $\theta$ we obtain $\overline{p_\Psi}(R)$, which is shown as color shading in Fig.~\ref{fig:theocal22}\,a) for $n=22$. The spectral line profile is then given by~\cite{stanojevic2008}
\begin{equation}\label{eq:spec}
    s(E)=\frac{\omega_\textrm{atom}^4}{(E_{pp}-E)^2}\int_0^\infty\sum_\Psi G(E-E_\Psi(R))\overline{p_\Psi}(R) R^2 \textrm{d}R \hspace{0.3cm},
\end{equation}
where $E=2 h \nu$, $E_{pp}$ is the energy of the $np_{3/2}np_{3/2}$ pair state, $\omega_\textrm{atom}$ is the atomic Rabi frequency, and $G$ is the spectral-density function of the laser, assumed to be a Gaussian with a full width at half maximum of 140~MHz. The resulting spectrum for $n=22$ is shown in Fig.~\ref{fig:theocal22}\,b) and in Fig.~\ref{fig:allspec} for $22\le n\le32$. An analysis of the interaction matrix shows that the coupling relevant for the excitation of $sf$ pair states is indeed the direct off-resonant dipole-quadrupole coupling to the $p_{3/2}p_{3/2}$ states.

The simulations reveal a small shift of the $sf$ resonances from the asymptotic pair energies (see Fig.~\ref{fig:theocal22}\,b)), which is also visible in the experimental spectra (Fig.~\ref{fig:allspec} and Fig.~\ref{fig:starkspec22}), in contrast to the dipole-dipole-coupled pair states observed in previous works~\cite{farooqi2003,stanojevic2008}. The reason for the shift of the dipole-quadrupole resonances lies in the $R^{-6}$ dependence of the off-resonant dipole-dipole interaction (which dominates the $R$-dependent energy shifts of the $sf$ pair states in our model at large distances) and the $R^{-8}$ dependence of the off-resonant dipole-quadrupole interaction (which dominates the admixture of $pp$ character to the $sf$ pair states at large separations). At distances where the latter interaction leads to a significant excitation probability ($R<0.25\mu\textrm{m}$ for $n=22$), the energy shifts caused by the former interactions are already significant.

The relative intensities of the dipole-dipole coupled $ss'$ features (marked by asterisks for $30\le n\le 32$ in Fig.~\ref{fig:allspec}) and the dipole-quadrupole coupled $sf$ features are well captured by our model, justifying the equal weighting of $L$-changing and $L$-conserving interaction terms (see discussion above). The dependence of the visibility of the pair resonances on the principal quantum number $n$ is also well captured by the model, although we did not investigate this dependence systematically in our experiments. Combining the relative electronic transition strengths from our model with Monte-Carlo simulations of the atom distribution in the ground state sample we can estimate the relative strength of pair resonances and atomic transitions. For the $23s_{1/2}18f_J$ resonance (see \textit{e.g.} Fig.~\ref{fig:starkspec22}) and the $22p_{3/2}$ atomic transition we find a ratio of about $10^{-4}$:$1$ which is consistent with our experimental observations, though the dynamic range of our intensity measurement prevents a quantitative assessment. The quadratic dependence of the pair excitation (see Eq.~\eqref{eq:spec}) on the laser intensity, which we also observe experimentally, together with the large detunings of these pair resonances (see Fig.~\ref{fig:allspec}) necessitate large atomic Rabi frequencies and thus the use of an intense laser for the excitation. However, at $n=23$, where the energy of the $23p_{3/2}23p_{3/2}$ pair state lies in between the energies of the $24s_{1/2}19f_{5/2}$ and $24s_{1/2}19f_{7/2}$ asymptotes, we expect that the molecular resonance should also be observable in spectra recorded with a (less intense) cw laser. At this $n$, an electric field around 10~V/cm would be sufficient to bring the $24s_{1/2}19f_{5/2}$ pair state into exact resonance with the $23p_{3/2}23p_{3/2}$ pair state, allowing for the investigation of F\"orster resonant energy transfer with a $1/R^4$ dependence of the coupling (instead of the well studied $1/R^3$ dependence~\cite{gallagher1982,gallagher2008}).

As discussed above, the dipole-quadrupole interaction couples electronic and rotational degrees of freedom, effectively entangling the electronic and rotational motion of the two atoms. Dipole-dipole interactions between Rydberg atoms have been suggested as means to manipulate their internuclear separation~\cite{wall2007}. In analogy, the dipole-quadrupole interaction might allow for the control of the rotational motion of two ultracold atoms. The significance of our results is  $(i)$ that they provide empirical evidence for the inadequacy of Eq.~\eqref{eq:Htot} and of g/u and electronic $\pm$ symmetry to describe, in general, interacting Rydberg atoms near pair resonances violating condition~\eqref{eq:parCond}, and $(ii)$ that they suggest and justify a simple procedure with which the rotational-electronic interactions for odd values of $L_\AR$+$L_\BR$ may be incorporated as effective potential terms in $V_\textrm{inter}$ when the rotational levels can be regarded as ``degenerate''. A similar assumption, though not explicitly declared as such, appears to have been made in the recent discussion of the permanent dipole moment of Rydberg atoms interacting with ground-state atoms~\cite{li2011}. The effective dipole-quadrupole terms in $V_\textrm{inter}$ give rise to potential-energy contributions scaling as $R^{-8}$ and it appears that whenever such dispersion terms are reported for homonuclear diatomic molecules (see \textit{e.g.} Ref.~\cite{dalgarno1967} for the $X^1\Sigma_g^+$ ground state of $H_2$) this assumption is also implicitly made.

\begin{acknowledgments}
We gratefully acknowledge discussions with Stephen L. Coy (MIT). This work is supported financially by the Swiss National Science Foundation under Project Nr.~200020-146759, the NCCR QSIT, and the EU Initial Training Network COHERENCE under grant FP7-PEOPLE-2010-ITN-265031.
\end{acknowledgments}


\begin{thebibliography}{36}%
\makeatletter
\providecommand \@ifxundefined [1]{%
 \@ifx{#1\undefined}
}%
\providecommand \@ifnum [1]{%
 \ifnum #1\expandafter \@firstoftwo
 \else \expandafter \@secondoftwo
 \fi
}%
\providecommand \@ifx [1]{%
 \ifx #1\expandafter \@firstoftwo
 \else \expandafter \@secondoftwo
 \fi
}%
\providecommand \natexlab [1]{#1}%
\providecommand \enquote  [1]{``#1''}%
\providecommand \bibnamefont  [1]{#1}%
\providecommand \bibfnamefont [1]{#1}%
\providecommand \citenamefont [1]{#1}%
\providecommand \href@noop [0]{\@secondoftwo}%
\providecommand \href [0]{\begingroup \@sanitize@url \@href}%
\providecommand \@href[1]{\@@startlink{#1}\@@href}%
\providecommand \@@href[1]{\endgroup#1\@@endlink}%
\providecommand \@sanitize@url [0]{\catcode `\\12\catcode `\$12\catcode
  `\&12\catcode `\#12\catcode `\^12\catcode `\_12\catcode `\%12\relax}%
\providecommand \@@startlink[1]{}%
\providecommand \@@endlink[0]{}%
\providecommand \url  [0]{\begingroup\@sanitize@url \@url }%
\providecommand \@url [1]{\endgroup\@href {#1}{\urlprefix }}%
\providecommand \urlprefix  [0]{URL }%
\providecommand \Eprint [0]{\href }%
\providecommand \doibase [0]{http://dx.doi.org/}%
\providecommand \selectlanguage [0]{\@gobble}%
\providecommand \bibinfo  [0]{\@secondoftwo}%
\providecommand \bibfield  [0]{\@secondoftwo}%
\providecommand \translation [1]{[#1]}%
\providecommand \BibitemOpen [0]{}%
\providecommand \bibitemStop [0]{}%
\providecommand \bibitemNoStop [0]{.\EOS\space}%
\providecommand \EOS [0]{\spacefactor3000\relax}%
\providecommand \BibitemShut  [1]{\csname bibitem#1\endcsname}%
\let\auto@bib@innerbib\@empty
%</preamble>
\bibitem [{\citenamefont {Robinson}\ \emph {et~al.}(2000)\citenamefont
  {Robinson}, \citenamefont {Laburthe-Tolra}, \citenamefont {Noel},
  \citenamefont {Gallagher},\ and\ \citenamefont {Pillet}}]{robinson2000}%
  \BibitemOpen
  \bibfield  {author} {\bibinfo {author} {\bibfnamefont {M.~P.}\ \bibnamefont
  {Robinson}}, \bibinfo {author} {\bibfnamefont {B.}~\bibnamefont
  {Laburthe-Tolra}}, \bibinfo {author} {\bibfnamefont {M.~W.}\ \bibnamefont
  {Noel}}, \bibinfo {author} {\bibfnamefont {T.~F.}\ \bibnamefont {Gallagher}},
  \ and\ \bibinfo {author} {\bibfnamefont {P.}~\bibnamefont {Pillet}},\ }\href
  {\doibase 10.1103/PhysRevLett.85.4466} {\bibfield  {journal} {\bibinfo
  {journal} {Phys. Rev. Lett.}\ }\textbf {\bibinfo {volume} {85}},\ \bibinfo
  {pages} {4466} (\bibinfo {year} {2000})}\BibitemShut {NoStop}%
\bibitem [{\citenamefont {Killian}\ \emph {et~al.}(2007)\citenamefont
  {Killian}, \citenamefont {Pattard}, \citenamefont {Pohl},\ and\ \citenamefont
  {Rost}}]{killian2007}%
  \BibitemOpen
  \bibfield  {author} {\bibinfo {author} {\bibfnamefont {T.~C.}\ \bibnamefont
  {Killian}}, \bibinfo {author} {\bibfnamefont {T.}~\bibnamefont {Pattard}},
  \bibinfo {author} {\bibfnamefont {T.}~\bibnamefont {Pohl}}, \ and\ \bibinfo
  {author} {\bibfnamefont {J.~M.}\ \bibnamefont {Rost}},\ }\href {\doibase
  10.1016/j.physrep.2007.04.007} {\bibfield  {journal} {\bibinfo  {journal}
  {Phys. Rep.}\ }\textbf {\bibinfo {volume} {449}},\ \bibinfo {pages} {77}
  (\bibinfo {year} {2007})}\BibitemShut {NoStop}%
\bibitem [{\citenamefont {Isenhower}\ \emph {et~al.}(2010)\citenamefont
  {Isenhower}, \citenamefont {Urban}, \citenamefont {Zhang}, \citenamefont
  {Gill}, \citenamefont {Henage}, \citenamefont {Johnson}, \citenamefont
  {Walker},\ and\ \citenamefont {Saffman}}]{isenhower2010}%
  \BibitemOpen
  \bibfield  {author} {\bibinfo {author} {\bibfnamefont {L.}~\bibnamefont
  {Isenhower}}, \bibinfo {author} {\bibfnamefont {E.}~\bibnamefont {Urban}},
  \bibinfo {author} {\bibfnamefont {X.~L.}\ \bibnamefont {Zhang}}, \bibinfo
  {author} {\bibfnamefont {A.~T.}\ \bibnamefont {Gill}}, \bibinfo {author}
  {\bibfnamefont {T.}~\bibnamefont {Henage}}, \bibinfo {author} {\bibfnamefont
  {T.~A.}\ \bibnamefont {Johnson}}, \bibinfo {author} {\bibfnamefont {T.~G.}\
  \bibnamefont {Walker}}, \ and\ \bibinfo {author} {\bibfnamefont
  {M.}~\bibnamefont {Saffman}},\ }\href {\doibase
  10.1103/PhysRevLett.104.010503} {\bibfield  {journal} {\bibinfo  {journal}
  {Phys. Rev. Lett.}\ }\textbf {\bibinfo {volume} {104}},\ \bibinfo {pages}
  {010503} (\bibinfo {year} {2010})}\BibitemShut {NoStop}%
\bibitem [{\citenamefont {Wilk}\ \emph {et~al.}(2010)\citenamefont {Wilk},
  \citenamefont {Ga\"etan}, \citenamefont {Evellin}, \citenamefont {Wolters},
  \citenamefont {Miroshnychenko}, \citenamefont {Grangier},\ and\ \citenamefont
  {Browaeys}}]{wilk2010}%
  \BibitemOpen
  \bibfield  {author} {\bibinfo {author} {\bibfnamefont {T.}~\bibnamefont
  {Wilk}}, \bibinfo {author} {\bibfnamefont {A.}~\bibnamefont {Ga\"etan}},
  \bibinfo {author} {\bibfnamefont {C.}~\bibnamefont {Evellin}}, \bibinfo
  {author} {\bibfnamefont {J.}~\bibnamefont {Wolters}}, \bibinfo {author}
  {\bibfnamefont {Y.}~\bibnamefont {Miroshnychenko}}, \bibinfo {author}
  {\bibfnamefont {P.}~\bibnamefont {Grangier}}, \ and\ \bibinfo {author}
  {\bibfnamefont {A.}~\bibnamefont {Browaeys}},\ }\href {\doibase
  10.1103/PhysRevLett.104.010502} {\bibfield  {journal} {\bibinfo  {journal}
  {Phys. Rev. Lett.}\ }\textbf {\bibinfo {volume} {104}},\ \bibinfo {pages}
  {010502} (\bibinfo {year} {2010})}\BibitemShut {NoStop}%
\bibitem [{\citenamefont {Farooqi}\ \emph {et~al.}(2003)\citenamefont
  {Farooqi}, \citenamefont {Tong}, \citenamefont {Krishnan}, \citenamefont
  {Stanojevic}, \citenamefont {Zhang}, \citenamefont {Ensher}, \citenamefont
  {Estrin}, \citenamefont {Boisseau}, \citenamefont {C\^ot\'e}, \citenamefont
  {Eyler},\ and\ \citenamefont {Gould}}]{farooqi2003}%
  \BibitemOpen
  \bibfield  {author} {\bibinfo {author} {\bibfnamefont {S.~M.}\ \bibnamefont
  {Farooqi}}, \bibinfo {author} {\bibfnamefont {D.}~\bibnamefont {Tong}},
  \bibinfo {author} {\bibfnamefont {S.}~\bibnamefont {Krishnan}}, \bibinfo
  {author} {\bibfnamefont {J.}~\bibnamefont {Stanojevic}}, \bibinfo {author}
  {\bibfnamefont {Y.~P.}\ \bibnamefont {Zhang}}, \bibinfo {author}
  {\bibfnamefont {J.~R.}\ \bibnamefont {Ensher}}, \bibinfo {author}
  {\bibfnamefont {A.~S.}\ \bibnamefont {Estrin}}, \bibinfo {author}
  {\bibfnamefont {C.}~\bibnamefont {Boisseau}}, \bibinfo {author}
  {\bibfnamefont {R.}~\bibnamefont {C\^ot\'e}}, \bibinfo {author}
  {\bibfnamefont {E.~E.}\ \bibnamefont {Eyler}}, \ and\ \bibinfo {author}
  {\bibfnamefont {P.~L.}\ \bibnamefont {Gould}},\ }\href {\doibase
  10.1103/PhysRevLett.91.183002} {\bibfield  {journal} {\bibinfo  {journal}
  {Phys. Rev. Lett.}\ }\textbf {\bibinfo {volume} {91}},\ \bibinfo {pages}
  {183002} (\bibinfo {year} {2003})}\BibitemShut {NoStop}%
\bibitem [{\citenamefont {Overstreet}\ \emph {et~al.}(2009)\citenamefont
  {Overstreet}, \citenamefont {Schwettmann}, \citenamefont {Tallant},
  \citenamefont {Booth},\ and\ \citenamefont {Shaffer}}]{overstreet2009}%
  \BibitemOpen
  \bibfield  {author} {\bibinfo {author} {\bibfnamefont {K.~R.}\ \bibnamefont
  {Overstreet}}, \bibinfo {author} {\bibfnamefont {A.}~\bibnamefont
  {Schwettmann}}, \bibinfo {author} {\bibfnamefont {J.}~\bibnamefont
  {Tallant}}, \bibinfo {author} {\bibfnamefont {D.}~\bibnamefont {Booth}}, \
  and\ \bibinfo {author} {\bibfnamefont {J.~P.}\ \bibnamefont {Shaffer}},\
  }\href {\doibase 10.1038/nphys1307} {\bibfield  {journal} {\bibinfo
  {journal} {Nat. Phys.}\ }\textbf {\bibinfo {volume} {5}},\ \bibinfo {pages}
  {581} (\bibinfo {year} {2009})}\BibitemShut {NoStop}%
\bibitem [{\citenamefont {Boisseau}\ \emph {et~al.}(2002)\citenamefont
  {Boisseau}, \citenamefont {Simbotin},\ and\ \citenamefont
  {C\^ot\'e}}]{boisseau2002}%
  \BibitemOpen
  \bibfield  {author} {\bibinfo {author} {\bibfnamefont {C.}~\bibnamefont
  {Boisseau}}, \bibinfo {author} {\bibfnamefont {I.}~\bibnamefont {Simbotin}},
  \ and\ \bibinfo {author} {\bibfnamefont {R.}~\bibnamefont {C\^ot\'e}},\
  }\href {\doibase 10.1103/PhysRevLett.88.133004} {\bibfield  {journal}
  {\bibinfo  {journal} {Phys. Rev. Lett.}\ }\textbf {\bibinfo {volume} {88}},\
  \bibinfo {pages} {133004} (\bibinfo {year} {2002})}\BibitemShut {NoStop}%
\bibitem [{\citenamefont {Stanojevic}\ \emph {et~al.}(2008)\citenamefont
  {Stanojevic}, \citenamefont {C\^ot\'e}, \citenamefont {Tong}, \citenamefont
  {Eyler},\ and\ \citenamefont {Gould}}]{stanojevic2008}%
  \BibitemOpen
  \bibfield  {author} {\bibinfo {author} {\bibfnamefont {J.}~\bibnamefont
  {Stanojevic}}, \bibinfo {author} {\bibfnamefont {R.}~\bibnamefont
  {C\^ot\'e}}, \bibinfo {author} {\bibfnamefont {D.}~\bibnamefont {Tong}},
  \bibinfo {author} {\bibfnamefont {E.~E.}\ \bibnamefont {Eyler}}, \ and\
  \bibinfo {author} {\bibfnamefont {P.~L.}\ \bibnamefont {Gould}},\ }\href
  {\doibase 10.1103/PhysRevA.78.052709} {\bibfield  {journal} {\bibinfo
  {journal} {Phys. Rev. A}\ }\textbf {\bibinfo {volume} {78}},\ \bibinfo
  {pages} {052709} (\bibinfo {year} {2008})}\BibitemShut {NoStop}%
\bibitem [{\citenamefont {Schwettmann}\ \emph {et~al.}(2006)\citenamefont
  {Schwettmann}, \citenamefont {Crawford}, \citenamefont {Overstreet},\ and\
  \citenamefont {Shaffer}}]{schwettmann2006}%
  \BibitemOpen
  \bibfield  {author} {\bibinfo {author} {\bibfnamefont {A.}~\bibnamefont
  {Schwettmann}}, \bibinfo {author} {\bibfnamefont {J.}~\bibnamefont
  {Crawford}}, \bibinfo {author} {\bibfnamefont {K.~R.}\ \bibnamefont
  {Overstreet}}, \ and\ \bibinfo {author} {\bibfnamefont {J.~P.}\ \bibnamefont
  {Shaffer}},\ }\href {\doibase 10.1103/PhysRevA.74.020701} {\bibfield
  {journal} {\bibinfo  {journal} {Phys. Rev. A}\ }\textbf {\bibinfo {volume}
  {74}},\ \bibinfo {pages} {020701} (\bibinfo {year} {2006})}\BibitemShut
  {NoStop}%
\bibitem [{\citenamefont {Walker}\ and\ \citenamefont
  {Saffman}(2008)}]{walker2008}%
  \BibitemOpen
  \bibfield  {author} {\bibinfo {author} {\bibfnamefont {T.~G.}\ \bibnamefont
  {Walker}}\ and\ \bibinfo {author} {\bibfnamefont {M.}~\bibnamefont
  {Saffman}},\ }\href {\doibase 10.1103/PhysRevA.77.032723} {\bibfield
  {journal} {\bibinfo  {journal} {Phys. Rev. A}\ }\textbf {\bibinfo {volume}
  {77}},\ \bibinfo {pages} {032723} (\bibinfo {year} {2008})}\BibitemShut
  {NoStop}%
\bibitem [{\citenamefont {Hougen}(1970)}]{hougen1970}%
  \BibitemOpen
  \bibfield  {author} {\bibinfo {author} {\bibfnamefont {J.}~\bibnamefont
  {Hougen}},\ }\href@noop {} {\emph {\bibinfo {title} {{The calculation of
  rotational energy levels and rotational line intensities in diatomic
  molecules}}}},\ \bibinfo {type} {{Report 115 }}\ (\bibinfo  {institution}
  {{Nat. Bur. Stand., Washington, DC, USA}},\ \bibinfo {year}
  {{1970}})\BibitemShut {NoStop}%
\bibitem [{\citenamefont {Lefebvre-Brion}\ and\ \citenamefont
  {Field}(2004)}]{lefebvre2004}%
  \BibitemOpen
  \bibfield  {author} {\bibinfo {author} {\bibfnamefont {H.}~\bibnamefont
  {Lefebvre-Brion}}\ and\ \bibinfo {author} {\bibfnamefont {R.~W.}\
  \bibnamefont {Field}},\ }in\ \href
  {http://www.sciencedirect.com/science/article/pii/B9780124414556500068}
  {\emph {\bibinfo {booktitle} {The Spectra and Dynamics of Diatomic
  Molecules}}}\ (\bibinfo  {publisher} {Academic Press},\ \bibinfo {address}
  {San Diego},\ \bibinfo {year} {2004})\ pp.\ \bibinfo {pages}
  {87--231}\BibitemShut {NoStop}%
\bibitem [{\citenamefont {Quack}(2011)}]{quack2011}%
  \BibitemOpen
  \bibfield  {author} {\bibinfo {author} {\bibfnamefont {M.}~\bibnamefont
  {Quack}},\ }\enquote {\bibinfo {title} {Fundamental symmetries and symmetry
  violations from high resolution spectroscopy},}\ in\ \href
  {http://onlinelibrary.wiley.com/doi/10.1002/9780470749593.hrs077/abstract}
  {\emph {\bibinfo {booktitle} {Handbook of High-resolution Spectroscopy}}},\
  \bibinfo {editor} {edited by\ \bibinfo {editor} {\bibfnamefont
  {M.}~\bibnamefont {Quack}}\ and\ \bibinfo {editor} {\bibfnamefont
  {F.}~\bibnamefont {Merkt}}}\ (\bibinfo  {publisher} {John Wiley \& Sons,
  Ltd},\ \bibinfo {year} {2011})\BibitemShut {NoStop}%
\bibitem [{\citenamefont {Hirschfelder}\ and\ \citenamefont
  {Meath}(1967)}]{hirschfelder1967}%
  \BibitemOpen
  \bibfield  {author} {\bibinfo {author} {\bibfnamefont {J.~O.}\ \bibnamefont
  {Hirschfelder}}\ and\ \bibinfo {author} {\bibfnamefont {W.~J.}\ \bibnamefont
  {Meath}},\ }\href
  {http://onlinelibrary.wiley.com/doi/10.1002/9780470143582.ch1/summary}
  {\bibfield  {journal} {\bibinfo  {journal} {Adv. Chem. Phys.}\ }\textbf
  {\bibinfo {volume} {12}},\ \bibinfo {pages} {3} (\bibinfo {year}
  {1967})}\BibitemShut {NoStop}%
\bibitem [{\citenamefont {Dalgarno}(1967)}]{dalgarno1967}%
  \BibitemOpen
  \bibfield  {author} {\bibinfo {author} {\bibfnamefont {A.}~\bibnamefont
  {Dalgarno}},\ }\href
  {http://onlinelibrary.wiley.com/doi/10.1002/9780470143582.ch3/summary}
  {\bibfield  {journal} {\bibinfo  {journal} {Adv. Chem. Phys.}\ }\textbf
  {\bibinfo {volume} {12}},\ \bibinfo {pages} {143} (\bibinfo {year}
  {1967})}\BibitemShut {NoStop}%
\bibitem [{\citenamefont {Flannery}\ \emph {et~al.}(2005)\citenamefont
  {Flannery}, \citenamefont {Vrinceanu},\ and\ \citenamefont
  {Ostrovsky}}]{flannery2005}%
  \BibitemOpen
  \bibfield  {author} {\bibinfo {author} {\bibfnamefont {M.~R.}\ \bibnamefont
  {Flannery}}, \bibinfo {author} {\bibfnamefont {D.}~\bibnamefont {Vrinceanu}},
  \ and\ \bibinfo {author} {\bibfnamefont {V.~N.}\ \bibnamefont {Ostrovsky}},\
  }\href {\doibase 10.1088/0953-4075/38/2/020} {\bibfield  {journal} {\bibinfo
  {journal} {J. Phys. B}\ }\textbf {\bibinfo {volume} {38}},\ \bibinfo {pages}
  {S279} (\bibinfo {year} {2005})}\BibitemShut {NoStop}%
\bibitem [{\citenamefont {Overstreet}\ \emph {et~al.}(2007)\citenamefont
  {Overstreet}, \citenamefont {Schwettmann}, \citenamefont {Tallant},\ and\
  \citenamefont {Shaffer}}]{overstreet2007}%
  \BibitemOpen
  \bibfield  {author} {\bibinfo {author} {\bibfnamefont {K.~R.}\ \bibnamefont
  {Overstreet}}, \bibinfo {author} {\bibfnamefont {A.}~\bibnamefont
  {Schwettmann}}, \bibinfo {author} {\bibfnamefont {J.}~\bibnamefont
  {Tallant}}, \ and\ \bibinfo {author} {\bibfnamefont {J.~P.}\ \bibnamefont
  {Shaffer}},\ }\href {\doibase 10.1103/PhysRevA.76.011403} {\bibfield
  {journal} {\bibinfo  {journal} {Phys. Rev. A}\ }\textbf {\bibinfo {volume}
  {76}},\ \bibinfo {pages} {011403} (\bibinfo {year} {2007})}\BibitemShut
  {NoStop}%
\bibitem [{\citenamefont {Fontana}(1961)}]{fontana1961}%
  \BibitemOpen
  \bibfield  {author} {\bibinfo {author} {\bibfnamefont {P.~R.}\ \bibnamefont
  {Fontana}},\ }\href {\doibase 10.1103/PhysRev.123.1865} {\bibfield  {journal}
  {\bibinfo  {journal} {Phys. Rev.}\ }\textbf {\bibinfo {volume} {123}},\
  \bibinfo {pages} {1865} (\bibinfo {year} {1961})}\BibitemShut {NoStop}%
\bibitem [{Note1()}]{Note1}%
  \BibitemOpen
  \bibinfo {note} {The $Y_{L\Omega }$ functions transform as $(-1)^{L-\Omega }$
  under $\sigma _v$~\cite {hougen1970}, regardless of the position of their
  center along the internuclear axis.}\BibitemShut {Stop}%
\bibitem [{\citenamefont {Salour}(1977)}]{salour1977}%
  \BibitemOpen
  \bibfield  {author} {\bibinfo {author} {\bibfnamefont {M.~M.}\ \bibnamefont
  {Salour}},\ }\href {\doibase 10.1016/0030-4018(77)90019-0} {\bibfield
  {journal} {\bibinfo  {journal} {Opt. Comm.}\ }\textbf {\bibinfo {volume}
  {22}},\ \bibinfo {pages} {202} (\bibinfo {year} {1977})}\BibitemShut
  {NoStop}%
\bibitem [{\citenamefont {Sa{\ss}mannshausen}\ \emph
  {et~al.}(2013)\citenamefont {Sa{\ss}mannshausen}, \citenamefont {Merkt},\
  and\ \citenamefont {Deiglmayr}}]{sasmannshausen2013}%
  \BibitemOpen
  \bibfield  {author} {\bibinfo {author} {\bibfnamefont {H.}~\bibnamefont
  {Sa{\ss}mannshausen}}, \bibinfo {author} {\bibfnamefont {F.}~\bibnamefont
  {Merkt}}, \ and\ \bibinfo {author} {\bibfnamefont {J.}~\bibnamefont
  {Deiglmayr}},\ }\href {\doibase 10.1103/PhysRevA.87.032519} {\bibfield
  {journal} {\bibinfo  {journal} {Phys. Rev. A}\ }\textbf {\bibinfo {volume}
  {87}},\ \bibinfo {pages} {032519} (\bibinfo {year} {2013})}\BibitemShut
  {NoStop}%
\bibitem [{\citenamefont {Zimmerman}\ \emph {et~al.}(1979)\citenamefont
  {Zimmerman}, \citenamefont {Littman}, \citenamefont {Kash},\ and\
  \citenamefont {Kleppner}}]{zimmerman1979}%
  \BibitemOpen
  \bibfield  {author} {\bibinfo {author} {\bibfnamefont {M.~L.}\ \bibnamefont
  {Zimmerman}}, \bibinfo {author} {\bibfnamefont {M.~G.}\ \bibnamefont
  {Littman}}, \bibinfo {author} {\bibfnamefont {M.~M.}\ \bibnamefont {Kash}}, \
  and\ \bibinfo {author} {\bibfnamefont {D.}~\bibnamefont {Kleppner}},\ }\href
  {\doibase 10.1103/PhysRevA.20.2251} {\bibfield  {journal} {\bibinfo
  {journal} {Phys. Rev. A}\ }\textbf {\bibinfo {volume} {20}},\ \bibinfo
  {pages} {2251} (\bibinfo {year} {1979})}\BibitemShut {NoStop}%
\bibitem [{\citenamefont {Stanojevic}\ \emph {et~al.}(2006)\citenamefont
  {Stanojevic}, \citenamefont {C\^ot\'e}, \citenamefont {Tong}, \citenamefont
  {Farooqi}, \citenamefont {Eyler},\ and\ \citenamefont
  {Gould}}]{stanojevic2006}%
  \BibitemOpen
  \bibfield  {author} {\bibinfo {author} {\bibfnamefont {J.}~\bibnamefont
  {Stanojevic}}, \bibinfo {author} {\bibfnamefont {R.}~\bibnamefont
  {C\^ot\'e}}, \bibinfo {author} {\bibfnamefont {D.}~\bibnamefont {Tong}},
  \bibinfo {author} {\bibfnamefont {S.~M.}\ \bibnamefont {Farooqi}}, \bibinfo
  {author} {\bibfnamefont {E.~E.}\ \bibnamefont {Eyler}}, \ and\ \bibinfo
  {author} {\bibfnamefont {P.~L.}\ \bibnamefont {Gould}},\ }\href {\doibase
  10.1140/epjd/e2006-00143-x} {\bibfield  {journal} {\bibinfo  {journal} {Eur.
  Phys. J. D}\ }\textbf {\bibinfo {volume} {40}},\ \bibinfo {pages} {3}
  (\bibinfo {year} {2006})}\BibitemShut {NoStop}%
\bibitem [{\citenamefont {Samboy}\ and\ \citenamefont
  {C\^ot\'e}(2011)}]{samboy2011}%
  \BibitemOpen
  \bibfield  {author} {\bibinfo {author} {\bibfnamefont {N.}~\bibnamefont
  {Samboy}}\ and\ \bibinfo {author} {\bibfnamefont {R.}~\bibnamefont
  {C\^ot\'e}},\ }\href {\doibase 10.1088/0953-4075/44/18/184006} {\bibfield
  {journal} {\bibinfo  {journal} {J. Phys. B}\ }\textbf {\bibinfo {volume}
  {44}},\ \bibinfo {pages} {184006} (\bibinfo {year} {2011})}\BibitemShut
  {NoStop}%
\bibitem [{\citenamefont {Le~Roy}(1973)}]{leroy1973}%
  \BibitemOpen
  \bibfield  {author} {\bibinfo {author} {\bibfnamefont {R.~J.}\ \bibnamefont
  {Le~Roy}},\ }in\ \href
  {http://pubs.rsc.org/en/content/chapter/bk9780851865065-00113/978-0-85186-506-5#!divabstract}
  {\emph {\bibinfo {booktitle} {Molecular Spectroscopy}}},\ Vol.~\bibinfo
  {volume} {1},\ \bibinfo {editor} {edited by\ \bibinfo {editor} {\bibfnamefont
  {R.~F.}\ \bibnamefont {Barrow}}, \bibinfo {editor} {\bibfnamefont {D.~A.}\
  \bibnamefont {Long}}, \ and\ \bibinfo {editor} {\bibfnamefont {D.~J.}\
  \bibnamefont {Millen}}}\ (\bibinfo  {publisher} {Royal Society of
  Chemistry},\ \bibinfo {address} {Cambridge},\ \bibinfo {year} {1973})\ pp.\
  \bibinfo {pages} {113--176}\BibitemShut {NoStop}%
\bibitem [{\citenamefont {Goy}\ \emph {et~al.}(1982)\citenamefont {Goy},
  \citenamefont {Raimond}, \citenamefont {Vitrant},\ and\ \citenamefont
  {Haroche}}]{goy1982}%
  \BibitemOpen
  \bibfield  {author} {\bibinfo {author} {\bibfnamefont {P.}~\bibnamefont
  {Goy}}, \bibinfo {author} {\bibfnamefont {J.~M.}\ \bibnamefont {Raimond}},
  \bibinfo {author} {\bibfnamefont {G.}~\bibnamefont {Vitrant}}, \ and\
  \bibinfo {author} {\bibfnamefont {S.}~\bibnamefont {Haroche}},\ }\href
  {\doibase 10.1103/PhysRevA.26.2733} {\bibfield  {journal} {\bibinfo
  {journal} {Phys. Rev. A}\ }\textbf {\bibinfo {volume} {26}},\ \bibinfo
  {pages} {2733} (\bibinfo {year} {1982})}\BibitemShut {NoStop}%
\bibitem [{\citenamefont {Marinescu}\ \emph {et~al.}(1994)\citenamefont
  {Marinescu}, \citenamefont {Sadeghpour},\ and\ \citenamefont
  {Dalgarno}}]{marinescu1994}%
  \BibitemOpen
  \bibfield  {author} {\bibinfo {author} {\bibfnamefont {M.}~\bibnamefont
  {Marinescu}}, \bibinfo {author} {\bibfnamefont {H.~R.}\ \bibnamefont
  {Sadeghpour}}, \ and\ \bibinfo {author} {\bibfnamefont {A.}~\bibnamefont
  {Dalgarno}},\ }\href {\doibase 10.1103/PhysRevA.49.982} {\bibfield  {journal}
  {\bibinfo  {journal} {Phys. Rev. A}\ }\textbf {\bibinfo {volume} {49}},\
  \bibinfo {pages} {982} (\bibinfo {year} {1994})}\BibitemShut {NoStop}%
\bibitem [{\citenamefont {Oka}(1974)}]{oka1974}%
  \BibitemOpen
  \bibfield  {author} {\bibinfo {author} {\bibfnamefont {T.}~\bibnamefont
  {Oka}},\ }in\ \href
  {http://www.sciencedirect.com/science/article/pii/S0065219908601153} {\emph
  {\bibinfo {booktitle} {Adv. At. Mol. Phys.}}},\ Vol.~\bibinfo {volume} {9},\
  \bibinfo {editor} {edited by\ \bibinfo {editor} {\bibnamefont {{{D.R.} Bates
  and {I.} Estermann}}}}\ (\bibinfo  {publisher} {Academic Press},\ \bibinfo
  {year} {1974})\ pp.\ \bibinfo {pages} {127--206}\BibitemShut {NoStop}%
\bibitem [{\citenamefont {Ovsiannikov}\ \emph {et~al.}(2011)\citenamefont
  {Ovsiannikov}, \citenamefont {Glukhov},\ and\ \citenamefont
  {Nekipelov}}]{ovsiannikov2011}%
  \BibitemOpen
  \bibfield  {author} {\bibinfo {author} {\bibfnamefont {V.~D.}\ \bibnamefont
  {Ovsiannikov}}, \bibinfo {author} {\bibfnamefont {I.~L.}\ \bibnamefont
  {Glukhov}}, \ and\ \bibinfo {author} {\bibfnamefont {E.~A.}\ \bibnamefont
  {Nekipelov}},\ }\href {\doibase 10.1088/0953-4075/44/19/195010} {\bibfield
  {journal} {\bibinfo  {journal} {J. Phys. B}\ }\textbf {\bibinfo {volume}
  {44}},\ \bibinfo {pages} {195010} (\bibinfo {year} {2011})}\BibitemShut
  {NoStop}%
\bibitem [{Note2()}]{Note2}%
  \BibitemOpen
  \bibinfo {note} {In order to reproduce the experimental spectrum around the
  asymptote $np\protect \tmspace +\thinmuskip {.1667em}np$ we found it
  necessary to include the pair states $n_{\protect \textrm {A}}s$-$n_{\protect
  \textrm {B}}s$, $n_{\protect \textrm {A}}p$-$n_{\protect \textrm {B}}p$,
  $n_{\protect \textrm {A}}d$-$n_{\protect \textrm {B}}d$, $n_{\protect \textrm
  {A}}f$-$n_{\protect \textrm {B}}f$, $n_{\protect \textrm {A}}s$-$n_{\protect
  \textrm {B}}f$, and $n_{\protect \textrm {A}}p$-$n_{\protect \textrm {B}}d$
  with $|n_{{\protect \textrm {A}}/{\protect \textrm {B}}}^* - n_p^*|\le 1$ and
  $|E_{n_{\protect \textrm {A}}\ell _{\protect \textrm {A}},n_{\protect \textrm
  {B}}\ell _{\protect \textrm {B}}}-E_{n_p pn_p p}|\le 0.6 |E_{n_p pn_p
  p}-E_{(n_p-1) p (n_p-1) p}|$ in the basis set.}\BibitemShut {Stop}%
\bibitem [{Note3()}]{Note3}%
  \BibitemOpen
  \bibinfo {note} {Because of a Cooper-minimum in the photoexcitation cross
  section from the 6$s_{1/2}$ state to the $np_{1/2}$ ($\epsilon p_{1/2}$)
  states (continuum) of Cs just above the ionization threshold~\cite
  {raimond1978}, only the photoexcitation to $np_{3/2}np_{3/2}$ states is of
  relevance.}\BibitemShut {Stop}%
\bibitem [{\citenamefont {Gallagher}\ \emph {et~al.}(1982)\citenamefont
  {Gallagher}, \citenamefont {Safinya}, \citenamefont {Gounand}, \citenamefont
  {Delpech}, \citenamefont {Sandner},\ and\ \citenamefont
  {Kachru}}]{gallagher1982}%
  \BibitemOpen
  \bibfield  {author} {\bibinfo {author} {\bibfnamefont {T.~F.}\ \bibnamefont
  {Gallagher}}, \bibinfo {author} {\bibfnamefont {K.~A.}\ \bibnamefont
  {Safinya}}, \bibinfo {author} {\bibfnamefont {F.}~\bibnamefont {Gounand}},
  \bibinfo {author} {\bibfnamefont {J.~F.}\ \bibnamefont {Delpech}}, \bibinfo
  {author} {\bibfnamefont {W.}~\bibnamefont {Sandner}}, \ and\ \bibinfo
  {author} {\bibfnamefont {R.}~\bibnamefont {Kachru}},\ }\href {\doibase
  10.1103/PhysRevA.25.1905} {\bibfield  {journal} {\bibinfo  {journal} {Phys.
  Rev. A}\ }\textbf {\bibinfo {volume} {25}},\ \bibinfo {pages} {1905}
  (\bibinfo {year} {1982})}\BibitemShut {NoStop}%
\bibitem [{\citenamefont {Gallagher}\ and\ \citenamefont
  {Pillet}(2008)}]{gallagher2008}%
  \BibitemOpen
  \bibfield  {author} {\bibinfo {author} {\bibfnamefont {T.~F.}\ \bibnamefont
  {Gallagher}}\ and\ \bibinfo {author} {\bibfnamefont {P.}~\bibnamefont
  {Pillet}},\ }in\ \href
  {http://www.sciencedirect.com/science/article/pii/S1049250X0800013X} {\emph
  {\bibinfo {booktitle} {Adv. At. Mol. Opt. Phys.}}},\ Vol.~\bibinfo {volume}
  {56},\ \bibinfo {editor} {edited by\ \bibinfo {editor} {\bibfnamefont
  {E.}~\bibnamefont {Arimondo}}}\ (\bibinfo  {publisher} {Academic Press},\
  \bibinfo {year} {2008})\ pp.\ \bibinfo {pages} {161--218}\BibitemShut
  {NoStop}%
\bibitem [{\citenamefont {Wall}\ \emph {et~al.}(2007)\citenamefont {Wall},
  \citenamefont {Robicheaux},\ and\ \citenamefont {Jones}}]{wall2007}%
  \BibitemOpen
  \bibfield  {author} {\bibinfo {author} {\bibfnamefont {M.~L.}\ \bibnamefont
  {Wall}}, \bibinfo {author} {\bibfnamefont {F.}~\bibnamefont {Robicheaux}}, \
  and\ \bibinfo {author} {\bibfnamefont {R.~R.}\ \bibnamefont {Jones}},\ }\href
  {\doibase 10.1088/0953-4075/40/18/009} {\bibfield  {journal} {\bibinfo
  {journal} {J. Phys. B}\ }\textbf {\bibinfo {volume} {40}},\ \bibinfo {pages}
  {3693} (\bibinfo {year} {2007})}\BibitemShut {NoStop}%
\bibitem [{\citenamefont {Li}\ \emph {et~al.}(2011)\citenamefont {Li},
  \citenamefont {Pohl}, \citenamefont {Rost}, \citenamefont {Rittenhouse},
  \citenamefont {Sadeghpour}, \citenamefont {Nipper}, \citenamefont {Butscher},
  \citenamefont {Balewski}, \citenamefont {Bendkowsky}, \citenamefont {Löw},\
  and\ \citenamefont {Pfau}}]{li2011}%
  \BibitemOpen
  \bibfield  {author} {\bibinfo {author} {\bibfnamefont {W.}~\bibnamefont
  {Li}}, \bibinfo {author} {\bibfnamefont {T.}~\bibnamefont {Pohl}}, \bibinfo
  {author} {\bibfnamefont {J.~M.}\ \bibnamefont {Rost}}, \bibinfo {author}
  {\bibfnamefont {S.~T.}\ \bibnamefont {Rittenhouse}}, \bibinfo {author}
  {\bibfnamefont {H.~R.}\ \bibnamefont {Sadeghpour}}, \bibinfo {author}
  {\bibfnamefont {J.}~\bibnamefont {Nipper}}, \bibinfo {author} {\bibfnamefont
  {B.}~\bibnamefont {Butscher}}, \bibinfo {author} {\bibfnamefont {J.~B.}\
  \bibnamefont {Balewski}}, \bibinfo {author} {\bibfnamefont {V.}~\bibnamefont
  {Bendkowsky}}, \bibinfo {author} {\bibfnamefont {R.}~\bibnamefont {Löw}}, \
  and\ \bibinfo {author} {\bibfnamefont {T.}~\bibnamefont {Pfau}},\ }\href
  {\doibase 10.1126/science.1211255} {\bibfield  {journal} {\bibinfo  {journal}
  {Science}\ }\textbf {\bibinfo {volume} {334}},\ \bibinfo {pages} {1110}
  (\bibinfo {year} {2011})}\BibitemShut {NoStop}%
\bibitem [{\citenamefont {Raimond}\ \emph {et~al.}(1978)\citenamefont
  {Raimond}, \citenamefont {Gross}, \citenamefont {Fabre}, \citenamefont
  {Haroche},\ and\ \citenamefont {Stroke}}]{raimond1978}%
  \BibitemOpen
  \bibfield  {author} {\bibinfo {author} {\bibfnamefont {J.~M.}\ \bibnamefont
  {Raimond}}, \bibinfo {author} {\bibfnamefont {M.}~\bibnamefont {Gross}},
  \bibinfo {author} {\bibfnamefont {C.}~\bibnamefont {Fabre}}, \bibinfo
  {author} {\bibfnamefont {S.}~\bibnamefont {Haroche}}, \ and\ \bibinfo
  {author} {\bibfnamefont {H.~H.}\ \bibnamefont {Stroke}},\ }\href {\doibase
  10.1088/0022-3700/11/24/004} {\bibfield  {journal} {\bibinfo  {journal} {J.
  Phys. B}\ }\textbf {\bibinfo {volume} {11}},\ \bibinfo {pages} {L765}
  (\bibinfo {year} {1978})}\BibitemShut {NoStop}%
\end{thebibliography}
\end{document}